# A MODEL-DRIVEN ENGINEERING APPROACH TO DEVELOP A COOPERATIVE INFORMATION SYSTEM

Mohamed Amroune[1], Pierre Jean Charrel[2], Nacereddine Zarour[1] and Jean Michel Inglebert[2]

[1]University of Constantine 2, Algeria
amroune@irit.fr, nasro-zarour@umc.edu
[2]University of Toulouse 2, France
charrel@univ-tlse2.fr, inglebert@iut-blagnac.fr

## ABSTRACT

*To reuse one or several existing systems in order to develop a complex system is a common practice in software engineering. This approach can be justified by the fact that it is often difficult for a single Information System (IS) to accomplish all the requested tasks. So, one solution is to combine many different ISs and make them collaborate in order to realize these tasks. We proposed an approach named AspeCiS (An Aspect-oriented Approach to Develop a Cooperative Information System) to develop a Cooperative Information System from existing ISs by using their artifacts such as existing requirements, and design. AspeCiS covers the three following phases: (i) discovery and analysis of Cooperative Requirements, (ii) design of Cooperative Requirements models, and (iii) preparation of the implementation phase. The main issue of AspeCiS is the definition of Cooperative Requirements using the Existing Requirements and Additional Requirements, which should be composed with Aspectual Requirements. We earlier studied how to elicit the Cooperative Requirements in AspeCiS (phase of discovery and analysis of Cooperative Requirements in AspeCiS) . We study here the second phase of AspeCiS (design of Cooperative Requirements models), by the way of a model weaving process. This process uses so-called AspeCiS Weaving Metamodel, and it weaves Existing and Additional Requirements models to realize Cooperative Requirements models.*

## 1. INTRODUCTION

Several reasons and economic factors lead organizations to interconnect their information systems in order to ensure a common global service and to support the tempo of change in the business world that is increasing at an exponential level. Consequently, they build a cooperative information system (CIS). We presented in [1] an approach named AspeCiS (An Aspect approach to develop Cooperative Information Systems), which develops Cooperative Information Systems by reusing the maximum of existing information systems artifacts. Furthermore, when a new requirement cannot be achieved directly by an existing IS, AspeCiS composes requirements issued from other ISs in order to fulfill this requirement. The main objectives of AspeCiS are: (i) to separate existing requirements from new requirements in the CIS; (ii) to provide a high degree of functional reuse, which helps to build again the same requirements on other existing ISs; (iii) to propose an aspect approach, which allows weaving existing and additional requirements on join points at the model level. AspeCiS includes three main phases which are: (i) discovery and analysis of Cooperative Requirements (CRs), (ii) conception of CRs models, and (iii) preparation of the implementation phase.

The Existing Requirements (ERs) are requirements provided by existing ISs. However, the Additional Requirements (ARs) are requirements that are not supported by any existing IS.

These kinds of requirements are used to define a set of CRs related to the CIS to be developed. In [1], we studied only how to elicit and analyze the CRs, but we have not developed how to define CRs models using existing models? So, in this paper, we try to answer this question. The basic assumption in model driven engineering (MDE) is to consider models as first class entities [2], [3],[4]. A model is an artifact that conforms to a metamodel so that it represents a given aspect of a system. Current MDE approaches usually have three representation levels for models: metametamodel, metamodel and terminal models [5].

In order to realize the potential benefits of models, AspeCiS must facilitates the specification and reuse of models, However, in AspeCiS, we use the models of ERs and ARs, and aspectual models, which should be composed with ERs and ARs models to produce the models of CRs. So, in AspeCis aspects models are those that are spread across and tangled with other design elements. The main objective of this paper is to propose a weaving models process in order to produce CRs's models.

The remainder of the paper is organized as follows: Section 2 presents an overview of AspeCiS. The AspeCiS Model Weaving process is detailed in section 3. Section 4 draws some examples. Section 5 provides some related works. Section 6 provides a summary of the paper and a brief overview of the continuation of this work.

# 2. ASPECIS: AN ASPECT APPROACH TO DEVELOP COOPERATIVE INFORMATION SYSTEM

## 2.1. The Form of Cooperation of AspeCiS

Today, enterprises operate in environments characterized by the increased competition, changes in customer demands, communications performance, etc. In this new context, and in order to cope with these business conditions, enterprises migrate to inter-organizational relationships [6], [7], [8] as a way to adapt to their new environment, gain competitive advantage, and, increase their efficiency. However, enterprise information systems were not designed to evolve with this new strategy of thinking. This adds another level of intricacy to an already complex IT infrastructure. For this reason, building effective enterprise cooperation is not an easy task, it requires a Cooperative Information System (CIS) to support this inter-enterprises cooperation.

According to our point of view, a CIS is a large scale information system that interconnects several systems of different organizations, sharing common objectives. We consider that the Model Driving Engineering (MDE) with the Aspect paradigm is one of the techniques for addressing these critical issues. Nowadays, MDE and Aspect Oriented Modeling (AOM) become among the most promising paradigm for leveraging enterprise information systems. It creates opportunities for enterprises to provide value added service. Thus, our research aims at developing a new approach called AspeCis, that ensures the effectiveness and efficiency of business cooperation based on the Aspect concept. AspeCiS focuses on three main issues that are respectively: Elicitation & analysis of requirements (called cooperative requirements) of the cooperative information system, which supports the business cooperation. Elaboration of cooperative requirements models, and transform cooperative requirements models into code.

In next section, we briefly excerpt the overview of AspeCiS from our previous work [1]. First, we define the following concepts; Existing Requirements (ERs), Additional requirements (ARs), Aspectual Requirements (AspRs) and Cooperative requirements (CRs).

## 2.2. The concept of Requirements in AspeCiS

Several definitions of requirement exist in the literature [9], but we adopt the following ones to differentiate between requirements in AspeCiS. So, in AspeCiS tree kind of requirements are defined.

**Existing Requirements (ERs).** They are statements of services or constraints provided by an existing system, which define how the system should react to particular inputs and how the system should behave in particular situations.
**Additional Requirements (ARs).** They are requirements which are not supported by any existing IS. In this case, other external information systems will be solicited to fulfill these ARs.
**Aspect Requirements (AspRs).** They are types of transverses requirements that must be woven with the existing requirements in order to reuse them.
**Cooperative Requirements (CRs).** They are goal requirements that will be refined to relate on ERs and eventually ARs, exhibiting what parts of existing systems requirements will be reused and composed, and what parts should be newly developed as shown in the metamodel of the **CRs** (see Figure 1).

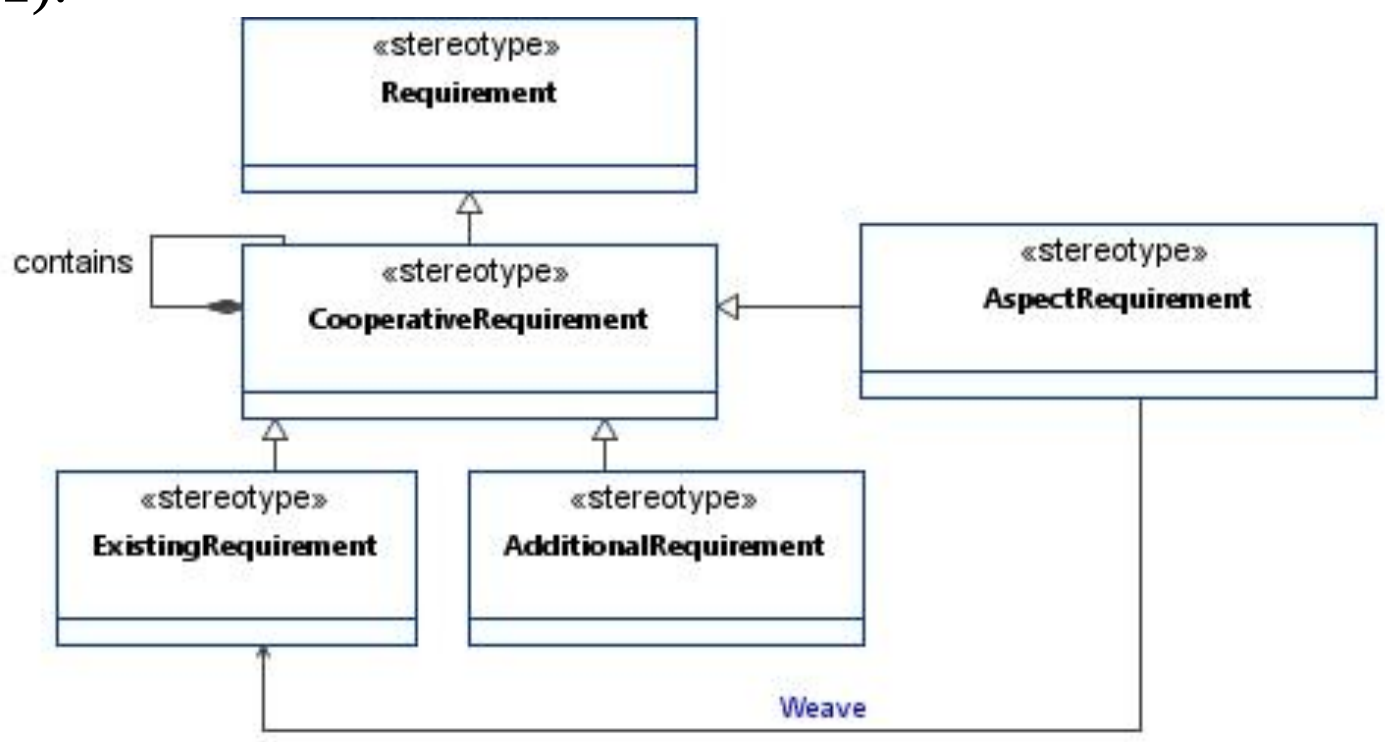

**Figure 1 A Cooperative requirement metamodel**

## 2.3.The Three Phases of AspeCiS

AspeCiS includes three main phases which are (cf. 2):
**Phase I: Elicitation and analysis of CRs.** This phase is composed of four steps which are: (1) the definition of CRs, (2) the refinement of CRs, (3) the formulation of CRs depending on the ERs and possibly with the definition of some ARs, (4) the selection of a set of Aspects-Requirement as can be seen in the figure 2.

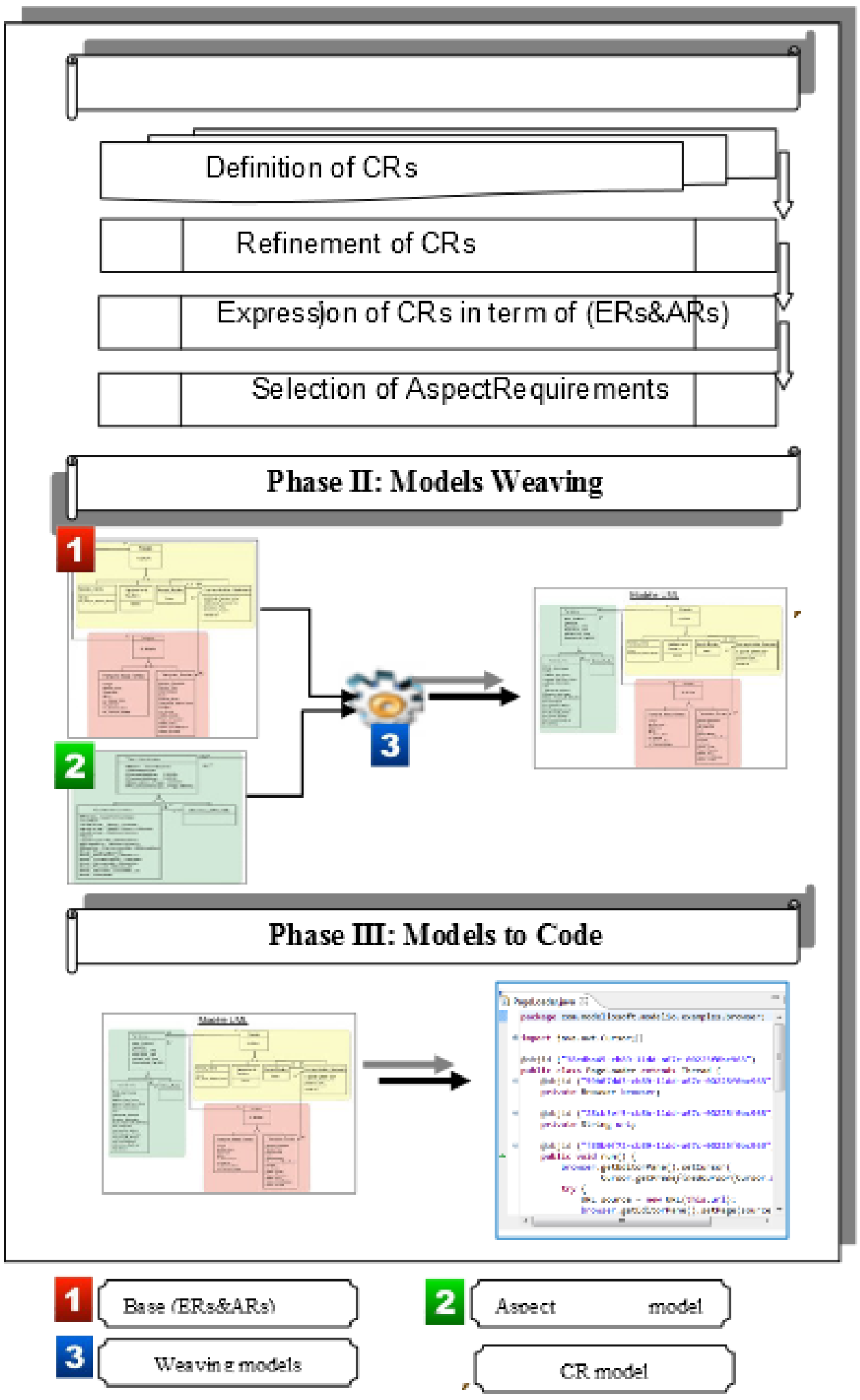

**Figure 2 Synopsis of AspeCiS approach**

**Phase II: Development of CRs models.** The Aspect Requirements models should be composed with ERs and ARs models to produce CRs models. This phase includes a conflict resolution task that can appear during the requirements composition process. We proposed in [10] a conflict resolution process among Aspect Requirements during the requirements engineering level: a priority value is computed for each Aspect Requirements, and it allows identifying a dominant Aspect Requirement on the basis of stakeholder priority. This process is more formal than those currently proposed, which requires a trade-off negotiation to resolve conflicts.

**Phase III: Preparing the implementation phase.** The purpose of this phase is to transform models into code templates.

### *2.3.1. Phase I: Definition of CRs*

The requirements elicitation activity offers means for the identification and analysis of all requirements. In the literature, several techniques for requirements elicitation are defined [9]. The choice of an elicitation technique depends on the time and resources available to the engineering requirements and, of course, the kind of information that needs to be elicited [9]. Nuseibeh presents an interesting classification of elicitation techniques [9]. This phase must be followed by a refinement process.

### *2.3.2. Phase I: Refinement of CRs*

The refinement process is composed by two actions, which are: the decomposition of CRs into a set of basic requirements, and the use of the inference relation.

***Decomposition of CRs.*** This action aims to decompose these requirements qualified as high-level requirements into a set of ERs and ARs (not decomposable). These requirements are connected by conjunctions or disjunctions nodes. We can distinguish ERs that can be used without any change in the definition of CRs and ERs that must be changed by means of appropriate Aspect Requirements.

C***ooperative Requirements Inference.*** Within the refinement phase, we can use the inference relation mentioned in [11]. The inference brings the following benefits: (i) It allows dealing with the redundancy of CRs. So, this relation avoids the definition of CRs that can be obtained simply by an inference relation. (ii) It allows the requirements engineer dealing with the problem of ambiguity that occurs when one CR has several possible interpretations. (iii) It allows reducing development cost and project schedule.

#### 2.2.3 Expression of CRs in terms of ERs and ARs

In this sub-phase we determine ERs and ARs involved in the definition of every CR which could not be inferred from others, i.e., we express CR using a combination of ERs and/or ARs.

#### *2.3.4. Phase I: Selection of the Aspect Requirements*

Usually, the reuse of ERs needs some modifications in order to reach CRs. These changes are assured by Aspect Requirement. A modification of an ER is the result of the weaving of a new behavior on this ER. This change is provided by the Aspect Requirement. So, we consider the Aspect Requirement as a specific kind of requirement which appear in the case of the definition of CRs. It must be woven with ERs in order to define CRs. An Aspect Requirement can be used several times in the definitions of the CRs. Thus, it is transverse.

## 3.AWMM: AN ASPECIS WEAVING METAMODEL

In this section, we present in detail the process, that we propose in this paper, to weave existing and additional models to produce models of CRs. The weaving concept is used to support such a decoupling among models. The weaving concept is not new and the definition of model weaving considered in this paper is an extension of the generic metamodel weaving proposed by Didonet Del Fabro et al. in [12].

The general operational context of this model weaving is depicted in Figure 3. It consists of the production of a weaving model WM representing the mapping between two metamodels: a left meatamodel LeftMM and a right metamodel RightMM. The WM model should be conform to a specific weaving metamodel WMM.

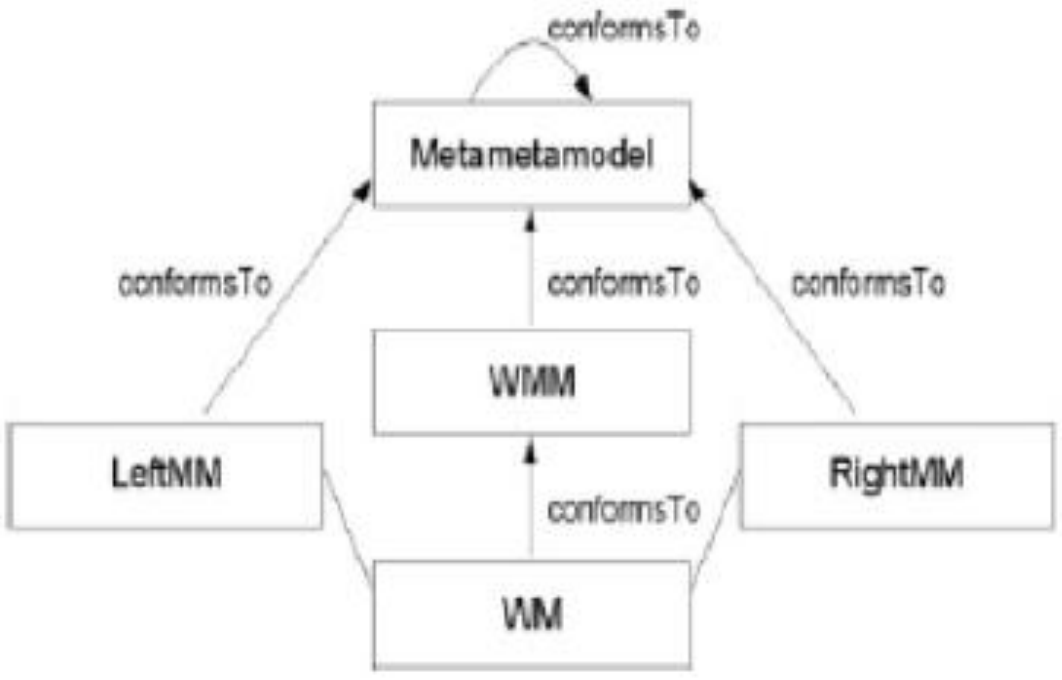

**Figure 3 AMW:Atlas Model Weaver metamodel. [12]**

The generic metamodel weaving proposed by Didonet Del Fabro et al. must be extended, to be used in our context. This extension (cf. figure 4), consists of the definition of a Core metamodel (AspeCiS-LeftMM), an Aspect Requirements metamodel (AspeCiS-RightMM) and a weaving metamodel called AWMM (for AspeCiS Weaving Metamodel) specific to our approach. So, the Core (Base) metamodel represents either an ERs or ARs metamodel. They are conformed to the same metamodel, which is the UML metamodel. We present these metamodels in the next sections.

We use weaving models to capture different kinds of links between model elements. The links have different semantics, depending on the application scenario. For instance (Attribute, Class) is a kind of link. It means that an attribute from Core model is added to a class from Aspect Requirement model. The semantic of links is not in the scope of this paper.

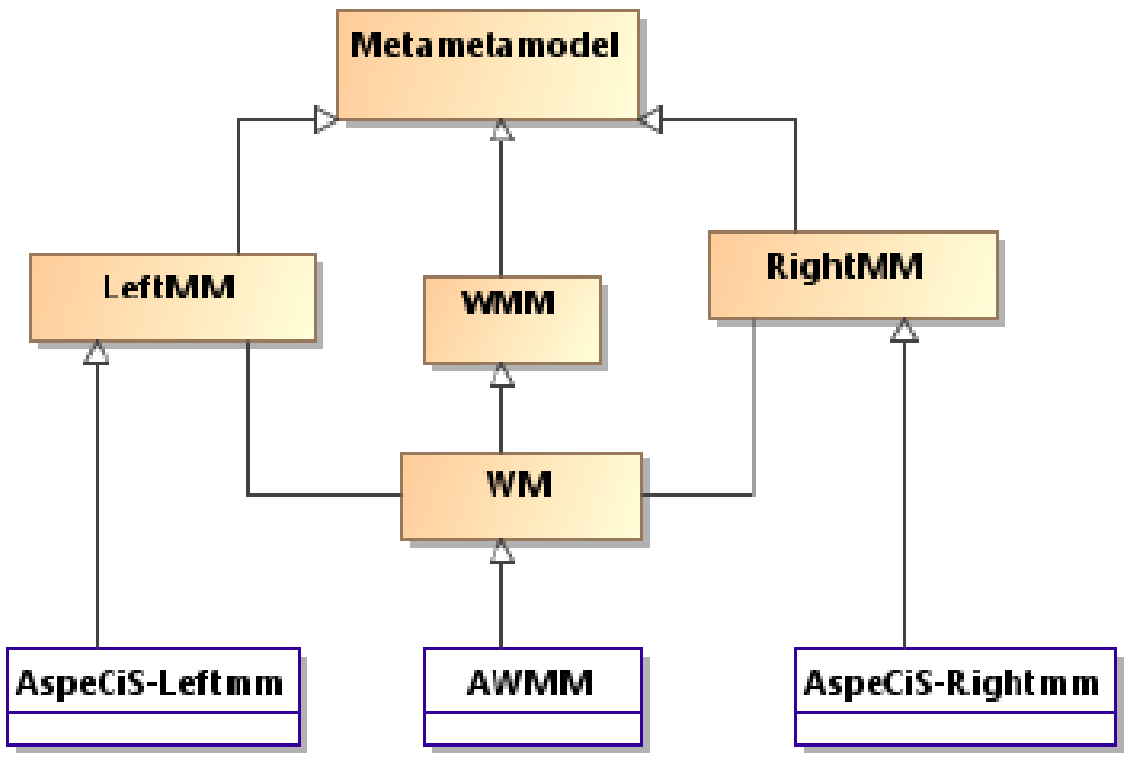

**Figure 4 AspeCiS Weaving Metamodel**

**3.1 AspeCiS-LeftMM: Existing and Additional Metamodel**. Each system can be described in structural and behavioral way. This paper focuses on the development of the structural dimension of a CIS. So, we use the class diagrams to represent all models. The Figure 5 describes the simplified UML class diagram metamodel used in the scope of this work. This metamodel can be used as an AspeCiS-LeftMM in order to describe ERs and ARs models. A Class contains features that may be either Attributes or Methods. A Class may be associated with several Associations. An Association is connected to the Classes by means of AssociationEnd elements that may be navigable or not. The AssociationClass element inherits from both the Association and the Class elements.

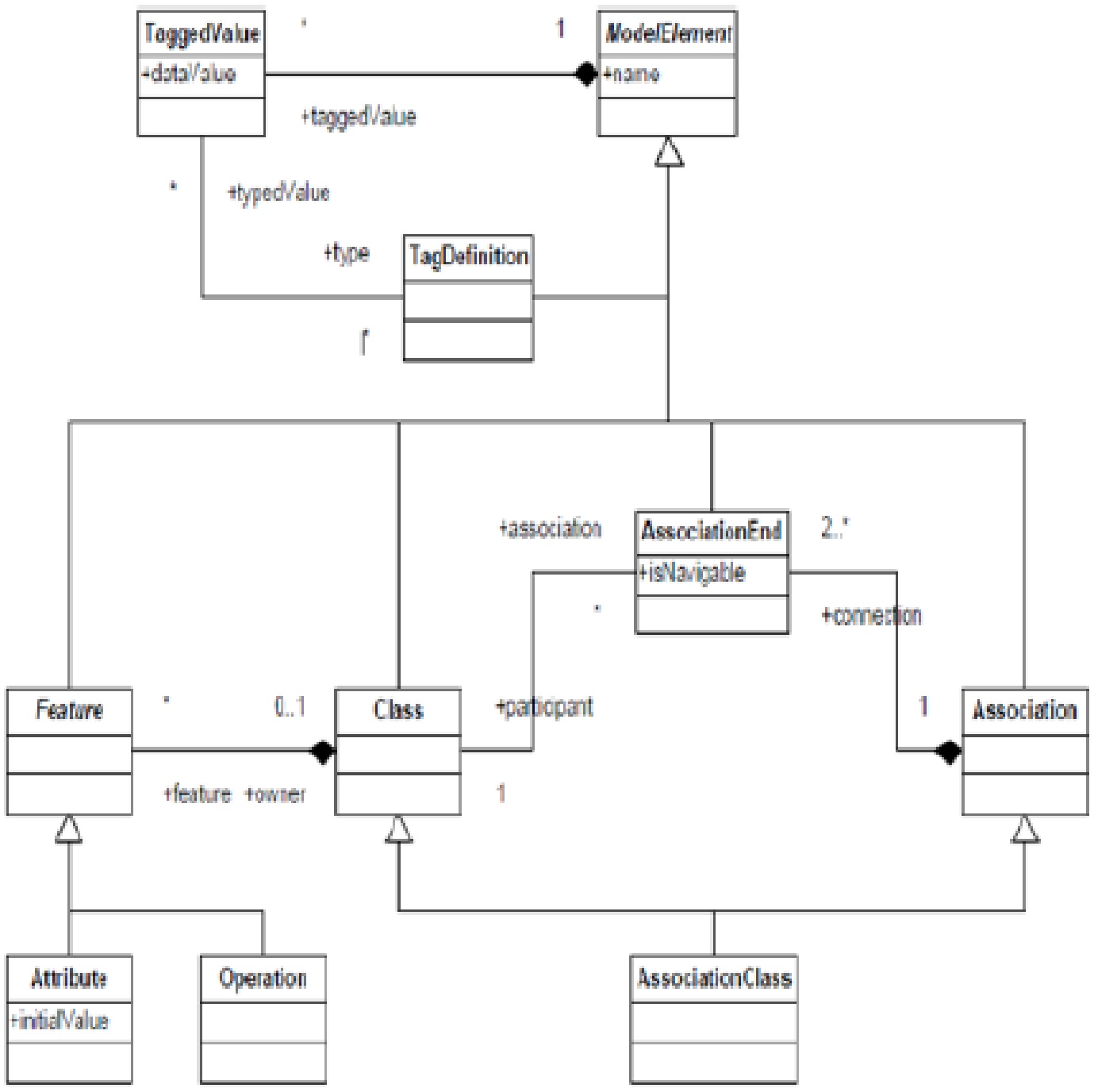

**Figure 5 AspeCiS-LeftMM metamodel**

## 3.2 AspeCiS-RightMM: Aspect Metamodel

An Aspect Requirement can be viewed as a modular way to implement crosscutting Requirements. In AspeCiS, the Aspect Requirement is very similar to a class. So, this similarity is also mentioned in [7], [5], [13], [19]. The Figure 6 illustrates the simplified metamodel of Aspect Requirement. However, the Aspect Requirement contains Pointcuts and Advices.

1) **Pointcuts**: The Pointcuts describe places where the Aspect Requirement should take effect. The Pointcuts define the Aspect Requirements structure. The stereotype Aspect Requirement extends the UML metaclass Class. The stereotype Pointcut extends the UML metaclass StructuralFeature (Figure 6).

2) **Advices:** The Advices define the Aspect Requirements behavior, (the modifications performed by the aspect). An Advice can be either Advice-AddElt to add an element, Advice-Update to modify an element, or an Advice-DeleteElt to remove an element. These elements can be attributes, classes or associations. Advices can be executed Before, After, of Pointcuts. This execution type is given in the attribute advicetype, this attribute is of type TypeAdvice, which is an enumeration containing the attributes Before and After. The stereotype Advice extends the UML metaclass BehavioralFeature (Figure 6). For each element (Class, Attribut and Association) to be modified a specific advice is defined.

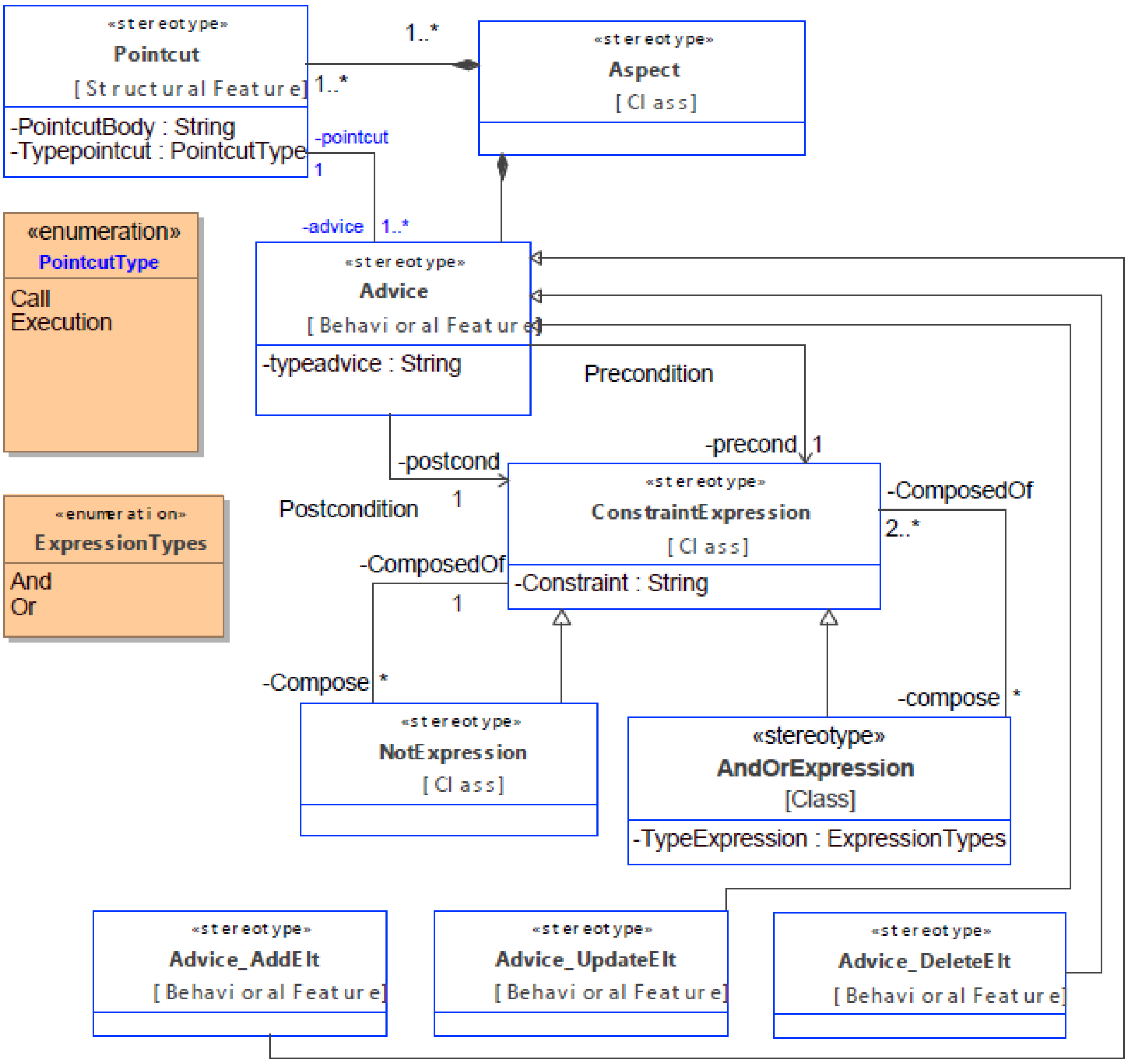

**Figure 6 The AspectRequirement Metamodel**

## 3.3. AWMM: AspeCiS Weaving Metamodel

Before presenting the AspeCiS metamodel, we briefly present the weaving metamodel (WM) proposed by Didonet Del Fabro et al. in [12]. This metamodel is illustrated in the Figure 7. This metamodel is composed by the following elements:

- ✓ WElement is the base element from which all other elements inherit. It has a name and a description.
- ✓ WModel represents the root element that contains all model elements. It is composed by the weaving elements and the references to woven models.
- ✓ WLink express a link between model elements, i.e., it has a simple linking semantics. To be able to express different link types and semantics, this element is extended with different metamodels.
- ✓ WLinkEnd represents a linked model element.
- ✓ WElementRef is associated with an identification function over the related elements. The function takes as parameter the model element to be linked and returns a unique identifier for this element.

WModelRef is similar to WElementRef element, but it references an entire model. The AspeCiS metamodel (AWMM) is an extension of the metamodel (WM) proposed by Didonet Del Fabro et al.

in [12]. AWMM is produced and depicted in Figure 7. So, we define two types of model weaving: WeavingCoreAspect and WeavingCoreAdditional. These weaving models are an extension of the WModel. Each weaving model is composed of the two models: Left and Right. The WeavingCoreAspect is used to weave an AspectOperatorRequirement model with CoreRequirementModel (Core models). Furthermore, the WeavingCoreAdditional is used to weave a CoreRequirementModel with an AdditionalRequirementModel. The result of the WeavingCoreAdditional can be used as an input to the WeavingCoreAspect.

### 3.3.1 Weaving Core and Aspect Models

This WeavingCoreAspect is composed of two models (CoreRequirementModel & AspectOperatorRequirement-Model). With respect to this metamodel, a WeavingCoreAspect consists of two models (CoreRequirementModel & AspectOperatorRequirement) related through weaving links (PointcutCoreAspect). The CoreRequirementModel (LeftModel) is an extension of WModelRef, it represents a model of ERs. The AspectOperatorRequirement( RightModel) represents a model of a Aspect Requirements. This model is also an extension of WModelRef. The WeavingCoreAspect is also composed of the Pointcut (PointcutCoreAspect), which is an extension of WLink. The PointcutCoreAspect is composed by two elements (left, right), these elements are an extension of WLinkEnd. The left element represents an artifact of CoreRequirementModel, and the right element represents an artifact of the Aspect Requirement.

### 3.3.2 Weaving Core and Additional Models

The CoreRequirementModel and AdditionalRequirementModel compose the WeavingCoreAdditional. They are an extension of WModelRef, the CoreRequirement-Model represents a model of ERs, and the AdditionalRequirementModel represents a model of ARs. The WeavingCoreAdditional is composed of the Pointcut PointutCoreAdd which is an extension of WLink. The PointcutCoreAdd is formed by two elements (leftCAdd, rightCAdd), these elements are an extension of WLinkEnd. The right element represents artifacts of AdditionalRequirementModel, and the left element represents artifacts of the CoreRequirementModel. These artifacts can represent an attribute, an association or a class.

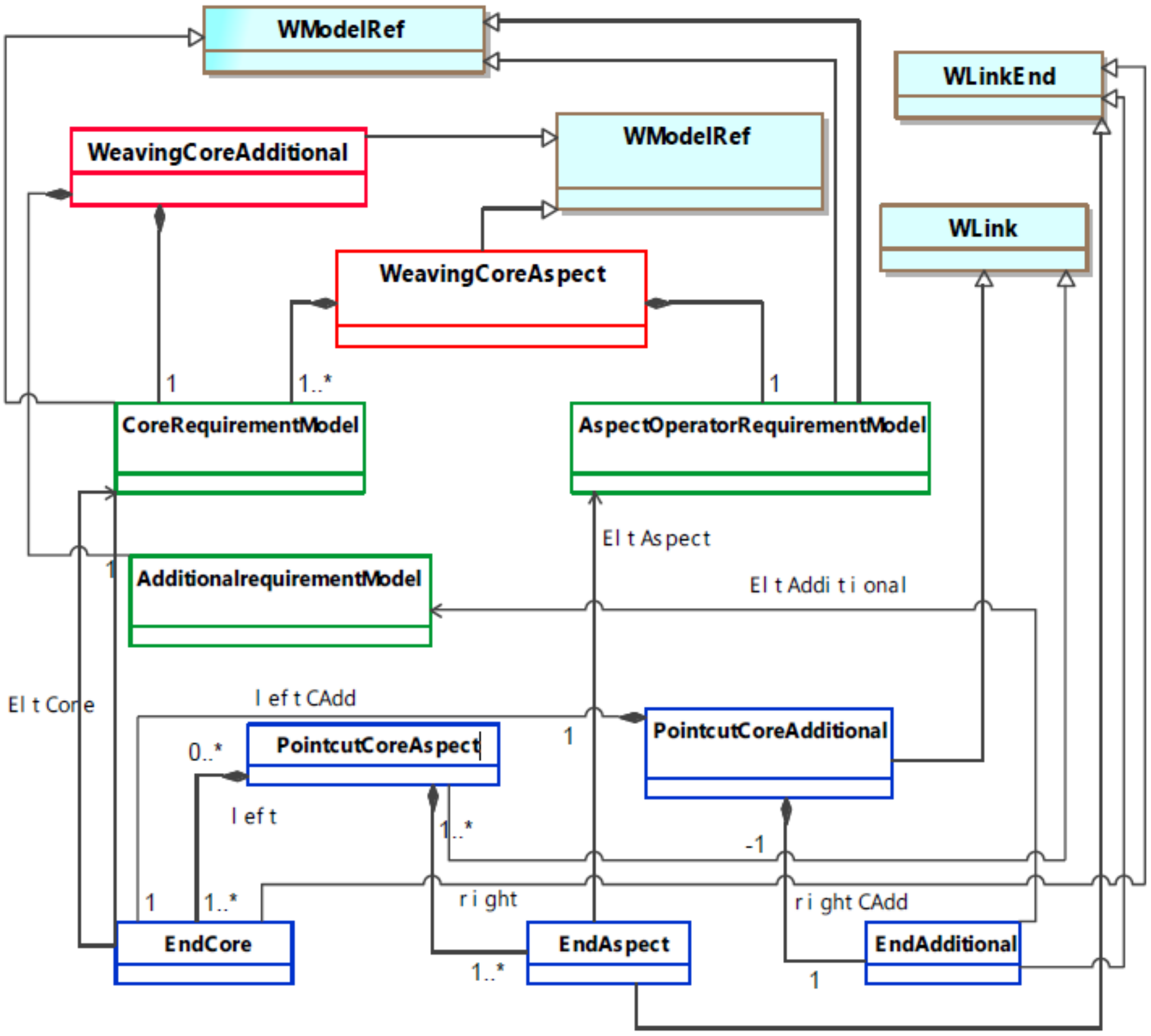

**Figure 7 AspeCiS Weaving Metamodel.**

## 4. The AWMM Usage

In the previous sections, we have presented a weaving process to produce CRs models. It is now convenient to better illustrate the use of this process.

### 4.1 Real Example

This example illustrates a part of the university student's management system. It consists of the management of the student's subscription in the High Graduate School in Algerian universities. The Hight Graduate School composes of several universities; it aims to assure a high formation of students. After completion of courses of study, the students receive a doctor's degree.

In this section, we illustrate how to weave two models (M1, M2) using AWMM, in order to produce a model of CRs. We intend to build a CIS able to manage a cooperative project involving several universities to provide High Graduate School.

Each university is supported by its existing IS. The new CIS is built on the basis of existing ISs (more details of this example are presented in [1]). At the requirements level of the existing ISs, the student subscription requirement is defined as:

ER1= "Every student may have a second subscription in the same university."

However, in the CIS, the CR is defined as:

CR1= "Every student can have a second subscription in the same university provided that the number of hours of the second specialty does not exceed 50% of the number of hours of the first one."

The existing ISs allow a second subscription in the same university. So, in order to participate in the Height Graduate School, each university must respect the constraint, of the number of hours for the second subscription, imposed by the Height Graduate School's regulation. This constraint is defined in the CR1 cited previously. Furthermore, the CIS to be developed, to support the management of this Height Graduate School, will be developed by reusing the existing ISs after some modifications.

At the model level of ISs, M1 models the ER1 (see figure 8). It represents the core (base) model that models the student subscription in the Height Graduate School. M1 is a class diagram conforms to UML class diagram metamodel, presented in the figure 5. M1 composes of three entities which are: University, Student and Speciality.

M2 (cf. figure 9) represents the Aspect Requirement model. M2 conforms to the Aspect Requirements metamodel presented in figure 6. It represents the aspect model. It also represents a constraint of the number of hours, which must be woven with M1 to define the CR1 model. In this example, the aspect model contains the advice advice_addElt, its role is to verify the number of hours before the call of the function Student.NewSpeciality(). This information is defined in the pointcut PointcutI through the (BodyAdvice) and the (Typepointcut= "call") (cf. figure 9).

## 4.2 THE AWMM IMPLEMENTATION

In order to validate the proposed weaving process, we have used the plug-in AMW in eclipse platform. The following code shows an extract of the AWMM in KM3 language. It contains two types of model weaving which are: WeavingCoreAspect and WeavingCoreAdditional. These weaving models are an extension of the WModel (lines (1),(2)).

```
package AWMM {
class WeavingCoreAspect       (1)
extends WModel { reference leftWCA container :
CoreRequirementModel;
reference rightWCA container:AspectOperatorRequirementModel;}
class CoreRequirementModel extends WModelRef {}
class AspectOperatorRequirementModel extends WModelRef {}
class PointcutCoreAspect extends WLink {
reference leftModel container : EndCore;
reference rightModel container : EndAspect ;
reference PointcutCoreAspect : PointcutCoreAspect;}
class WeavingCoreAdditional        (2)
extends WModel {
reference leftmodel container : CoreRequirementModel;
reference rightmodel container : AdditionalrequirementModel;}
class PointcutCoreAdditional extends WLink {
reference leftCAdd container :EndCore;
reference rightCAdd container :EndAdditional;
}
}
```

In this example, we use the WeavingCoreAspect to add two operations to M1, especially to the Student class. These operations are called Before the call of the Student.NewSubscription() operation, in order to add a second subscription. The first operation Verify Specialty.NbreOfHours(IdSpecialty) consists of the computation of the number of hours for the new subscription, the result of this operation is used by the second operation get SecondSpecialty() to verify the constraint imposed to authorize or not a second subscription. In this example, we use eclipse platform with plug-ins EMF and AMW in order to implement AWMM.

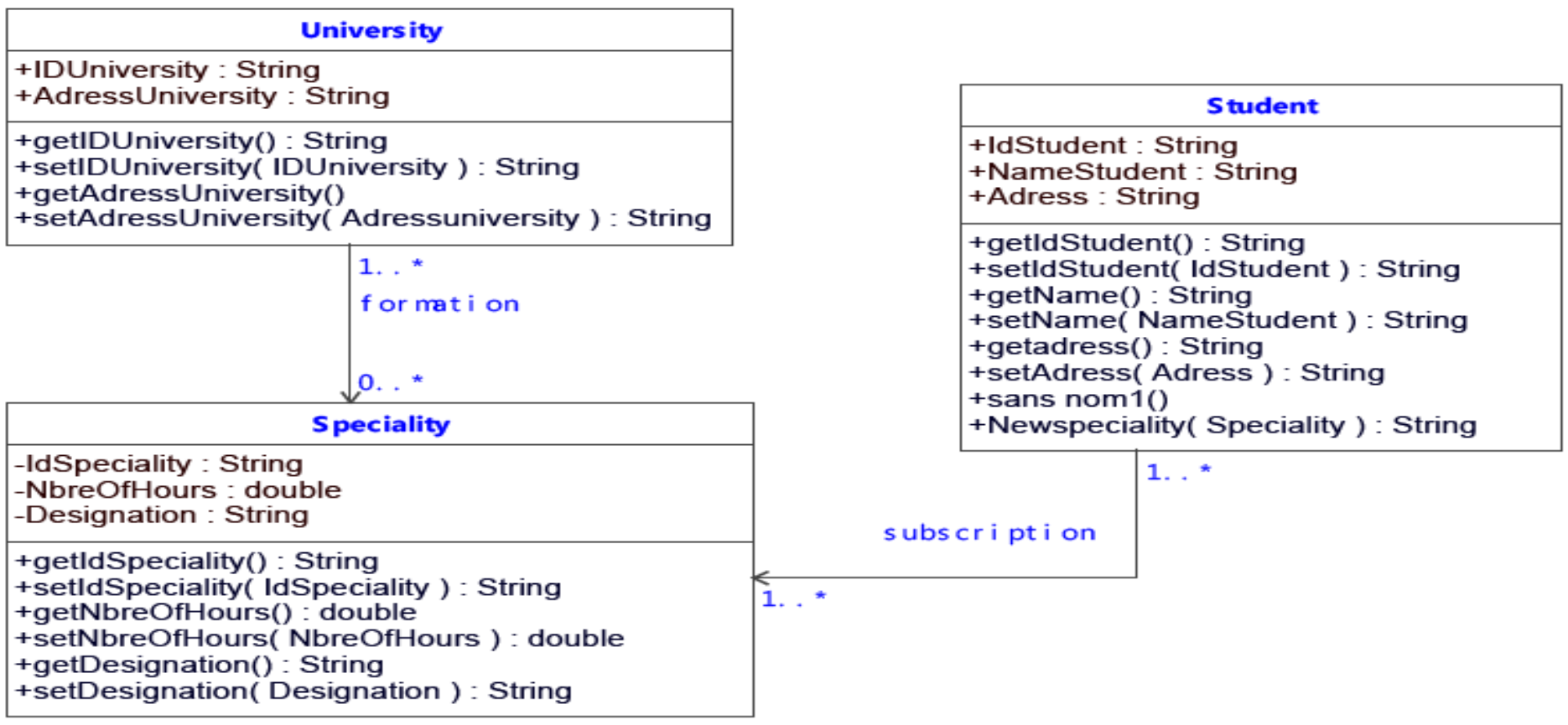

**Figure 8 M1: A Core model (Left)**

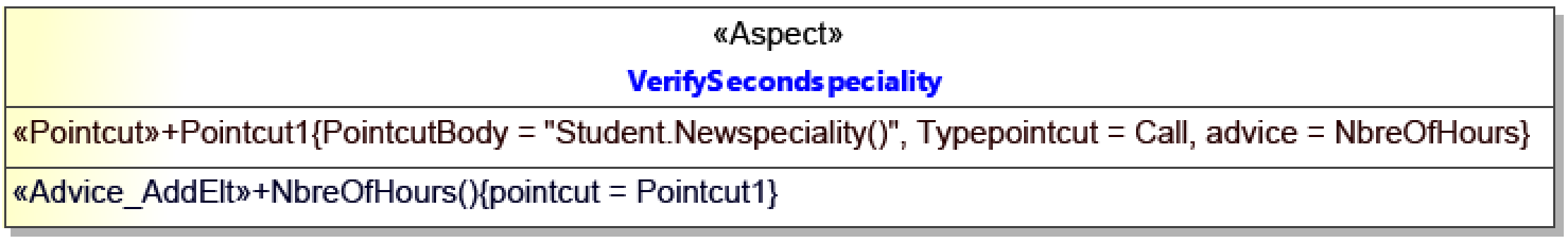

**Figure 9 An Aspect model (Right)**

## 5. RELATED WORKS

Over the last past years, several approaches to weave models have been proposed, [14], [15], [16], [11], [17], [18]. The database metadata integration and evolution is considered, as one of the important domains, where the model weaving is widely applied. In this context the work presented in [14] proposes a generic metamodel management algebra which uses algebraic operators to manage mappings and models. In [15], authors introduce a UML extension to express mappings between models using diagrams, and they illustrate how the extension can be used in metamodeling. The extension is inspired by mathematical relations and it is based upon ideas presented in [16]. This work defines transformation relationships between different components of a language definition, rendered as a metamodel. Another generic metamodel to support weaving operation is given in [11]: here, the approach is based on the possible extensibility and variability of mappings among metamodels and it is supported by a prototypical implementation.

The problem of heterogeneous composition is addressed in [17]: if several view models are expressed in different Domain Specific Modeling Languages and if these models are used to specify a system, then it is necessary to compose them to generate the application. To avoid this problem, this approach transforms the models into a same Metamodel. Therefore, when all the view models are transformed, in a same metamodel specified with a common language, it is possible to apply a homogeneous composition to obtain the final application. The approach proposed in [16], discusses a formal approach to define how the distinct concerns of a Web application can be better connected by means of weaving models. The main idea consists of the specification of weaving operators to establish relationships among the models that describe the various perspectives of the application being developed.

After reviewing, these approaches are not dedicated to the development of CIS. However, we can highlight, that it is possible, to use the model weaving techniques to reuse a set of existing models in order to produce cooperative models of a CIS. Therefore, our proposed weaving process aims to develop CIS by reusing models of existing ISs. The suggested weaving metamodel is an extension of the generic metamodel weaving proposed by Didonet Del Fabro et al [11]. This work aims to complete the development of the second phase of our approach called AspeCiS presented in [1].

## 6. CONCLUSION

It is frequently not easy for a sole Information System (IS) to achieve a new emerging complex task. One solution is to make existing ISs collaborate to realize this task. So, in systems engineering, reuse is defined as the utilization of previously developed systems engineering products or artifacts such as architectures and requirements.

In a previous work [1], we proposed an approach named AspeCiS to develop a Cooperative IS (CIS) from existing ISs by using their artifacts such as requirements, and design. AspeCiS describes four types of requirements: Cooperative Requirements (CR), Existing Requirements (ER), Additional Requirements (AR) and Aspect Requirements (AspRs). ERs are requirements provided by existing ISs, ARs are requirements that are not supported by any existing IS. Aspect Requirements are transverse requirements which appear in the definition of CRs.

AspeCiS contains three main phases, which are: discovery and analysis of CRs, design of CRs models, and preparation of the implementation phase. In the present work, we proposed a Model-Driven Engineering (MDE) approach to develop the CRs models of AspeCiS. So, to define CRs models we propose a MDE based weaving process which relies on a metamodel called AWMM, to be used to capture different kinds of links between input model elements.

The weaving process uses three proposed metamodels wich are: AspeCiSLeftMM, AspeCiS-RightMM and AWMM. The AWMM is an extension of the generic metamodel weaving (Atlas Model Weaver) proposed by Didonet Del Fabro et al [13]. We presented also an AspeCiS-LeftMM metamodel to represent existing, additional requirements and AspeCiS-RightMM metamodel to represent aspect requirements. Aspect Requirements must be woven with ERs and ARs in order to define the CRs related to the CIS to be developed. The weaving metamodel is used to capture different kinds of links between model elements. These links are represented by a set of PointCuts, which have different semantics. We will define these semantics in our future work.

## ACKNOWLEDGEMENTS

This research is partially supported by the PHC TASSILI project under the number 10MDU817. We thank the anonymous reviewers for providing valuable comments.

## REFERENCES

[1]. M. Amroune, J.-M. Inglebert, N. Zarour, and P.-J. Charrel, “Aspecis: An aspect-oriented approach to develop a cooperative information system,” in Model and Data Engineering - First International Conference, MEDI 2011, Obidos, Portugal, ser. Lecture Notes in Computer Science, vol. 6918. Springer, 2011, pp. 122–132.

[2]. R. B. France and B. Rumpe, “Model-driven development of complex software: A research roadmap,” in FOSE, 2007, pp. 37–54.

[3]. B. Selic, “The pragmatics of model-driven development,”IEEE Software, vol. 20, no. 5, pp. 19–25, 2003.

[4]. R. Silaghi, F. Fondement, and A. Strohmeier, “Towards an mda-oriented uml profile for distribution,” in EDOC, 2004, pp. 227–239.

[5]. A. Y. Halevy, Z. G. Ives, and A. Doan, Eds., Rondo: a programming platform for generic model management. ACM, 2003.

[6]. K. V. Andersen, J. K. Debenham, and R. Wagner, Eds., How to design a Loose Inter-Organizational Workflow: An illustrative case study, ser. Lecture Notes in Computer
[7]. Science, vol. 3588. Springer, 2005.
[8]. P. W. P. J. Grefen, H. Ludwig, A. Dan, and S. Angelov, "An analysis of web services support for dynamic business process outsourcing," Information & Software Technology, vol. 48, no. 11, pp. 1115–1134, 2006.
[9]. P. W. P. J. Grefen, N. Mehandjiev, G. Kouvas, G. Weichhart, and R. Eshuis, "Dynamic business network process management in instant virtual enterprises," Computers in Industry, vol. 60, no. 2, pp. 86–103, 2009.
[10]. B. Nuseibeh and S. Easterbrook, "Requirements engineering: a roadmap," in Proceedings of the Conference on The Future of Software Engineering, ser. ICSE '00. New York, NY, USA: ACM, 2000, pp. 35–46. [Online]. Available: http://doi.acm.org/10.1145/336512.336523
[11]. M. Amroune, J. M. Inglebert, N. Zarour, and P. J. Charrel, "Article: A conflict resolution process in aspecis approach," International Journal of Computer Applications, vol. 44, no. 10, pp. 14–21, April 2012, published by Foundation of Computer Science, New York, USA.
[12]. I. Jureta, A. Borgida, N. A. Ernst, and J. Mylopoulos, "Techne: Towards a new generation of requirements modeling languages with goals, preferences, and inconsistency handling," in RE, 2010, pp. 115–124.
[13]. M. D. Del Fabro, J. B´ezivin, F. Jouault, E. Breton, and G. Gueltas, "AMW: a Generic Model Weaver," in Procs. of IDM05, 2005.
[14]. M. R. Djabri and M. Amroune, "A uml profile for aspectc++," in Proceedings of International Conference on Information Technology and e-Services (ICITeS), 2012. Conference Publications IEEE, 2012, pp. 1–6.
[15]. S. Diehl, J. T. Stasko, and S. N. Spencer, Eds., Proceedings ACM 2003 Symposium on Software Visualization, San Diego, California, USA, June 11-13, 2003. ACM, 2003.
[16]. J.-M. J´ez´equel, H. Hußmann, and S. Cook, Eds., UML 2002 - The Unified Modeling Language, 5th International Conference, Dresden, Germany, September 30 – October 4,2002, Proceedings, ser. Lecture Notes in Computer Science, vol. 2460. Springer, 2002.
[17]. A. Cicchetti and D. D. Ruscio, "Decoupling web application concerns through weaving operations," Sci. Comput. Program., vol. 70, no. 1, pp. 62–86, 2008.
[18]. A. Yie, R. Casallas, D. Deridder, and D. Wagelaar, "A practical approach to multi-modeling views composition," ECEASST, vol. 21, 2009.
[19]. J. Evermann, "A meta-level specification and profile for aspectj in uml," Journal of Object Technology, vol. 6, no. 7, pp. 27–49, 2007.

**Authors**

**Mohamed Amroune** is currently a Ph.D. Student, at the University of Toulouse II, Mirail, France and University of Tebessa, Algeria. He received his Engineer and Magister degrees in Software Engineering and Artificial Intelligence & Data Bases from the USTHB University of Algiers , Algeria, and the University of Tebessa, Algeria, in 1993 and 2007, respectively. His research interests include Information System, Requirements Engineering, Cooperation and Aspect oriented software development.

**Pierre Jean Charrel** is a Professor at the Computer Sciences Department of University of Toulouse, France. His current research activities are conducted at the IRIT laboratory, University of Toulouse. He currently heads PHC CMEP Tassili project nr 10MDU817 with LIRE Laboratory of Mentouri University of Constantine, Algeria. His research interests include requirements engineering and knowledge engineering, in the context of cooperative information systems.

**Nacereddine Zarour** is a Professor at the Computer Sciences Department of University of Constantine 2, Algeria. His current research activities are conducted at the LIRE laboratory, University of Constantine. He heads the project of PHC CMEP Tassili with IRIT laboratory of Toulouse 2 University. His research interests include advanced information systems, particularly cooperative information systems, architectures (based on SOA, SMA, ..), and requirements engineering.

**Jean Michel Inglebert** is a Doctor at the Computer Sciences Department of University of Toulouse, France. His current research activities are conducted at the IRIT laboratory, University of Toulouse.